\documentclass[
showpacs,
floatfix,
aps,
prb,
amsmath,
twocolumn,
superscriptaddress, 
tightenlines
groupaddress,
]{revtex4}

\usepackage{graphicx}
\usepackage{dcolumn}
\usepackage{amsmath}
\usepackage{amsfonts}
\usepackage{bm}
\usepackage{epsfig}
\usepackage{times}
\usepackage[varg]{txfonts}

\newcommand{\be}{\begin{equation}}
\newcommand{\ee}{\end{equation}}
\newcommand{\bea}{\begin{eqnarray}}
\newcommand{\eea}{\end{eqnarray}}

\def\Edc{\mathcal{E}_{\mathrm{dc}}}

\def\edc{\epsilon_{\mathrm{dc}}}

\def\oc{\omega_{\mbox{\scriptsize {c}}}}

\def\rc{R_{\mbox{\scriptsize {c}}}}

\def\tauq{\tau_{\mbox{\scriptsize {q}}}}

\begin{document}

\title{
Zero differential resistance in two-dimensional electron systems at large filling factors 
}

\author{A.\,T. Hatke}
\affiliation{School of Physics and Astronomy, University of Minnesota, Minneapolis, Minnesota 55455, USA}

\author{H.\,-S. Chiang}
\affiliation{School of Physics and Astronomy, University of Minnesota, Minneapolis, Minnesota 55455, USA}

\author{M.\,A. Zudov}
\email[Corresponding author: ]{zudov@physics.umn.edu}
\affiliation{School of Physics and Astronomy, University of Minnesota, Minneapolis, Minnesota 55455, USA}

\author{L.\,N. Pfeiffer}
\affiliation{Princeton University, Department of Electrical Engineering, Princeton, NJ 08544, USA}

\author{K.\,W. West}
\affiliation{Princeton University, Department of Electrical Engineering, Princeton, NJ 08544, USA}

\received{10 February 2010}

\begin{abstract}
We report on a state characterized by a zero differential resistance observed in very high Landau levels of
a high-mobility two-dimensional electron system.
Emerging from a minimum of Hall field-induced resistance oscillations at low temperatures, this state exists over a continuous range of magnetic fields extending well below the onset of the Shubnikov-de Haas effect.
The minimum current required to support this state is largely independent on the magnetic field, while the maximum current increases with the magnetic field tracing the onset of inter-Landau level scattering.
\end{abstract}
\maketitle

Over the past decade it was realized that high mobility two-dimensional electron systems (2DESs) exhibit an array of fascinating phenomena occurring in very high Landau levels where the Shubnikov-de Haas oscillations (SdHOs) are not yet resolved. 
Among these are three classes of magneto-oscillations, namely microwave-,\citep{zudov:2001a,ye:2001,zudov:2004,willett:2004,mani:2004e,dorozhkin:2005,studenikin:2005,yuan:2006,studenikin:2007,wirthmann:2007,wiedmann:2008,wiedmann:2009,andreev:2009,hatke:2009a,tung:2009,fedorych:2010} phonon-,\citep{zudov:2001b,zhang:2008,bykov:2005b,hatke:2009b} and Hall field-\citep{yang:2002,bykov:2005c,zhang:2007a,hatke:2009c} induced resistance oscillations (HIROs).
Remarkably, the minima of microwave-induced oscillations can evolve into states with zero resistance.\citep{mani:2002,zudov:2003,yang:2003,smet:2005,mani:2005,zudov:2006b,yang:2006,bykov:2006}
These exotic states are currently understood in terms of the absolute negative resistance which leads to an instability with respect to formation of current domains.\citep{andreev:2003,anderson:2003,auerbach:2005,finkler:2009}
Unfortunately, direct experimental confirmation of the domain structure has proven difficult in irradiated 2DES and awaits future studies.
It is therefore of great interest to explore if other classes of oscillations give rise to phenomenologically similar states.
Recently, experiments revealed states with zero differential resistance which emerged from the maxima of microwave-induced resistance oscillations \citep{zhang:2007c,hatke:2008} and from the maxima of the SdHOs.\citep{bykov:2007,romero:2008,zhang:2009,vitkalov:2009}
Such states are analogous to the radiation-induced zero-resistance states in a sense that they can also be explained by the domain model.\citep{bykov:2007}

In this Rapid Communication we report on another state characterized by a zero differential resistance which require neither microwave irradiation nor the Shubnikov-de Haas effect.
This state emerges from a minimum of HIROs in a high mobility 2DES at low temperatures.
Appearing in very high Landau levels, this state is observed over a continuous magnetic field range extending well below the onset of the SdHOs.
The minimum current required to support such a state is largely independent on the magnetic field, while the maximum current increases roughly linearly with the magnetic field tracing the onset of inter-Landau level scattering.
According to the domain model,\citep{bykov:2007} these currents should be associated with currents inside the domains.

The data presented in this Rapid Communication were obtained on a Hall bar (width $w=100$ $\mu$m) etched from a symmetrically doped GaAs/AlGaAs quantum well.
After a brief low-temperature illumination with visible light, density and mobility were $n_e\simeq 3.8\times 10^{11}$ cm$^{-2}$ and $\mu\simeq 1.0 \times 10^{7}$ cm$^{2}$/Vs, respectively.
Differential resistivity, $r_{xx}\equiv dV_{xx}/dI$, was measured using a quasi-dc (a few hertz) lock-in technique at temperatures ranging from $T = 1.5$ to $3.0$ K.

Since the zero differential resistance state (ZdRS) reported here originates from the minimum of HIROs, we first discuss the basic physical picture behind this effect.
According to the ``displacement'' model,\citep{vavilov:2007,lei:2007,khodas:2008,khodas:2010} HIROs originate from the impurity-mediated transitions between Landau levels tilted by the Hall electric field, $\Edc=\rho_{H}j$, where $\rho_H$ is the Hall resistivity and $j=I/w$ is the current density.
In this scenario, a dominant scattering process involves an electron which is  backscattered off an impurity.
The guiding center of such an electron is displaced by a distance equal to the cyclotron diameter $2\rc$.
When $2\rc$ matches an integral multiple of the real-space Landau level separation, the probability of such events is enhanced.
This enhancement manifests as a maximum in the differential resistivity occurring whenever $\edc\equiv e\Edc(2\rc)/\hbar\oc$ ($\oc$ is the cyclotron frequency) is equal to an integer.\citep{zhang:2007a,vavilov:2007}
As we will show, disappearance of the ZdRS is directly related to the fundamental HIRO peak at $\edc=1$.

At $2\pi\edc \lesssim 1$, the theory\citep{vavilov:2007} predicts another source of non-linearities known as the ``inelastic''\citep{dmitriev:2005} mechanism.
In this model a dc field creates a nonequilibrium distribution of electron states which, in turn, leads to a resistance drop.
We note that both models were developed in the limit of strongly overlapped Landau levels, a condition which is not always satisfied in experiments.\citep{zhang:2007a,zhang:2007b,bykov:2007,romero:2008}
As a result, the relative importance of these mechanisms at $2\pi\edc \lesssim 1$ remains poorly understood and calls for further investigations.
At the same time this regime is directly relevant to the formation of states with zero differential resistance, as we show below.

%%%%%%%%%%%%%%%%%%%%%%%%%%%%%%%%%%%%%%%%%%%%%%%%%
%fig 1
\begin{figure}[h]
\includegraphics{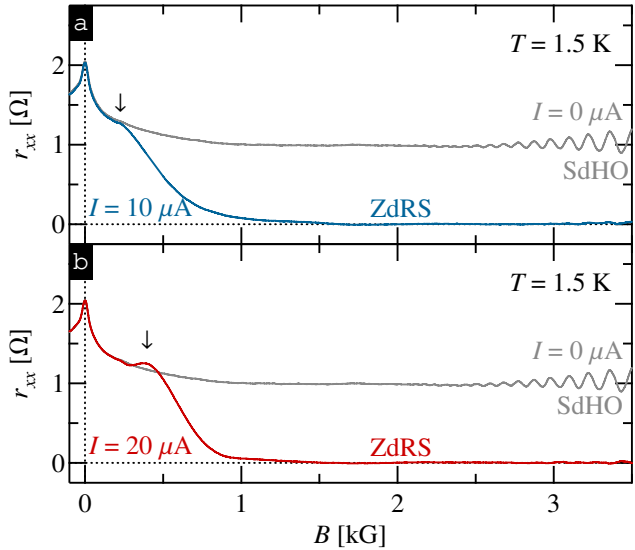}
\vspace{-0.1 in}
\caption{(Color online)
(a)\,[(b)] Differential magnetoresistivity $r_{xx}(B)$ measured at $I=10$ $\mu $A [$20$ $\mu$A] and $T \simeq 1.5$ K.
Magnetoresistivity $\rho_{xx}(B)$ at $I=0$ is shown for comparison. 
}
\vspace{-0.1 in}
\label{f1}
\end{figure}
%%%%%%%%%%%%%%%%%%%%%%%%%%%%%%%%%%%%%%%%%%%%%%%%%
We now present our experimental results. 
In Fig.\,\ref{f1} (a) and 1 (b) we plot the longitudinal differential magnetoresistivity $r_{xx}(B)$ acquired at $T = 1.5$ K and currents of $I=10$ $\mu$A and $I=20$ $\mu$A, respectively.
For comparison, each panel also includes the linear response ($I=0$) longitudinal magnetoresistivity $\rho_{xx}(B)$, which is essentially featureless and exhibits only the SdHOs starting to develop at $B\gtrsim 2.5$ kG.
At low magnetic fields, the 2DES remains in the linear response regime as manifested by the overlapping curves obtained at zero and finite currents.
However at higher magnetic fields, the data obtained at finite currents show several distinct characteristics signaling strong nonlinearities.
First, the differential resistivity at 10 and 20 $\mu$A reveals a pronounced peak at $B\simeq 0.2$ kG and $B\simeq 0.4$ kG, respectively (cf.,\,$\downarrow$).
This peak occurs at $\edc \simeq 1$ and, as discussed above, originates from the resonantly enhanced scattering due to electron transitions between neighboring Hall field-tilted Landau levels.
However, the most remarkable feature of Fig.\,\ref{f1} is the dramatic drop of $r_{xx}$ at higher $B$ (smaller $\edc$) which extends all the way to zero at $B \gtrsim 1$ kG. 
This drop marks a transition to the ZdRS whose generic characteristics are the focus of this Rapid Communication. 

We now show that the ZdRS presented in Fig.\,\ref{f1} is qualitatively different from those reported earlier.\cite{zhang:2007c,bykov:2007}
Indeed, the ZdRS in Ref.\,37 emerged from a maximum of microwave-induced resistance oscillations whereas our experiments are performed without microwaves.
The ZdRS reported in Ref.\,37 occurred in high magnetic fields where the linear response resistivity is dominated by the SdHOs, and the ZdRS were formed at the discrete values of the magnetic field corresponding to the SdHO maxima (odd filling factors).
Clearly, the ZdRS shown in Fig.\,{\ref{f1} extends over a continuous range of magnetic fields and does not rely on the existence of the SdHOs at all; similar to the microwave-induced zero-resistance states, it persists to magnetic fields much lower than the onset of the SdHOs.
Finally, we note that our data reveal neither negative spikes in differential resistance preceding the ZdRS nor temporal fluctuations reported in Ref.\,39.

%%%%%%%%%%%%%%%%%%%%%%%%%%%%%%%%%%%%%%%%%%%%%%%%%
\begin{figure}[t]
\includegraphics{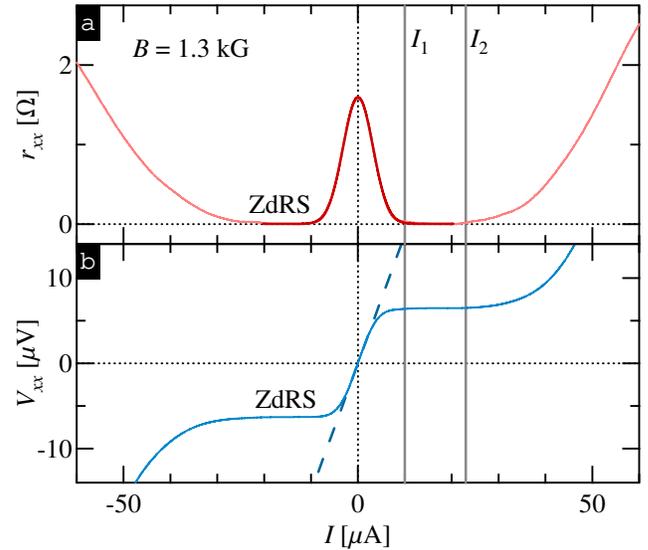}
\vspace{-0.1 in}
\caption{(Color online)
(a) Differential magnetoresistivity $r_{xx}(I) $ and (b) voltage $V_{xx} $ at $B=1.3$ kG and $T \simeq 1.5$ K. 
Dashed line in (b) represents Ohm's law which holds at small $I$.
Vertical lines mark the ZdRS critical currents, $I_1$ and $I_2$.
}
\vspace{-0.1 in}
\label{f2}
\end{figure}
%%%%%%%%%%%%%%%%%%%%%%%%%%%%%%%%%%%%%%%%%%%%%%%%%

In the context of the domain model, the range of currents supporting the ZdRS is of particular interest.
To investigate this range it is convenient to employ an alternative measurement technique in which the magnetic field $B$ is held constant and the current $I$ is varied.
This approach readily reveals both the minimum and the maximum currents for a given magnetic field. 
One example of such a measurement performed at $B=1.3$ kG  is presented in Fig.\,\ref{f2}(a) showing the differential resistivity $r_{xx}$ as a function of applied current $I$.
We observe that $r_{xx}$ exhibits a dramatic drop with increasing current which eventually evolves into a state with zero differential resistance (cf.,\,``ZdRS'').
Formation of the ZdRS can therefore be characterized by a current $I_1\simeq 10$ $\mu$A (cf., left line).
Once formed, the ZdRS persists up to a current $I_2\simeq 23$ $\mu$A (cf., right line) above which the differential resistivity starts to increase.

The drop in the $r_{xx}$ preceding the ZdRS can be examined quantitatively by fitting the data with a Gaussian $r_{xx}(I)=r_{xx}(0)\exp(-I^2/\Delta_1^2)$.
Here $r_{xx}(0)$ is the linear response resistivity and $\Delta_1$ is the characteristic current which can be related to $I_1$.
An example of such a fit over the current range from $-20$ to $+20$ $\mu $A is shown in Fig.\,\ref{f2}\,(a) by a dark line.
It describes the experimental data remarkably well yielding $\Delta_1 \simeq 4.5$ $\mu$A from which $I_1$ can be estimated as $I_1 \simeq 2\Delta_1$.

The phenomenon can also be illustrated by a current-voltage characteristic, shown in Fig.\,\ref{f2}\,(b), which is obtained by integrating the data shown in Fig.\,\ref{f2}\,(a).
Concurrent with the drop in the $r_{xx}$ observed in Fig.\,\ref{f2}\,(a), the longitudinal voltage $V_{xx}$ departs from Ohm's law (cf.,\,dashed line) and saturates to a plateau which extends over a finite current range.
Within this range (cf.,\, vertical lines), the voltage is independent of the applied current and, as we show next, is also largely insensitive to the magnetic field.
%%%%%%%%%%%%%%%%%%%%%%%%%%%%%%%%%%%%%%%%%%%%%%%%%
\begin{figure}[t]
\includegraphics{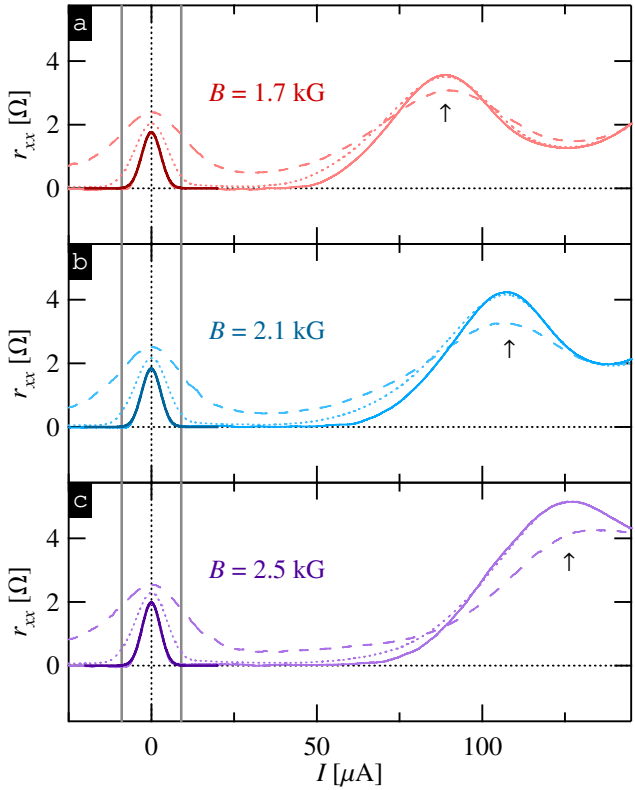}
\vspace{-0.1 in}
\caption{(Color online)
Differential resistivity $r_{xx}$ vs $I$ at (a) 1.7 kG\,, (b) 2.1 kG\,, and (c) 2.5 kG\ at $T = 1.5$ K (solid), 2.0 K (dotted), and 3.0 K (dashed). 
$I_H$ is marked by $\uparrow$ (see text).
}
\vspace{-0.1 in}
\label{f3}
\end{figure}
%%%%%%%%%%%%%%%%%%%%%%%%%%%%%%%%%%%%%%%%%%%%%%%%%

In Fig.\,\ref{f3}\,(a)-3(c) we present the differential resistivity $r_{xx}(I)$ obtained at higher magnetic fields, i.e. (a) $B = 1.7$ kG, (b) 2.1 kG, and (c) 2.5 kG, each measured at three different temperatures $T = 1.5$ K (solid line), 2.0 K (dotted line), and 3.0 K (dashed line).
At $T=1.5$ K all data show well developed ZdRS.
At $T=2.0$ K the ZdRS becomes narrower and at $T=3.0$ K are totally destroyed.
While recent experiments suggest electron-electron interactions as the origin of HIROs temperature dependence,\cite{hatke:2009c} this issue has not yet been theoretically considered\cite{vavilov:2007,lei:2007} and awaits future studies.  
In what follows we thus limit our discussion to the $T=1.5$ K data showing well developed ZdRS.

Examination of the data in Fig.\,\ref{f3} reveals that the linear response resistivity $r_{xx}(0)$ and the lower critical current (cf.,\,vertical lines), $I_{1}$, both have very weak dependence on the magnetic field.
As a result, the voltage at the plateau, which is equal to the area under the zero bias peak, is also largely independent on $B$.
As illustrated in Fig.\,\ref{f3}\,(a)-3(c) all the $T=1.5$ K data at $I\lesssim I_1$ are well described by $\exp(-I^2/\Delta_1^2)$ (cf., dark lines) with $\Delta_1 \simeq 4.0 $ $\mu $A.
We note that this value is slightly lower than the one obtained at $B=1.3$ kG and that the ZdRS is not developed in our 2DES at $B\lesssim 1$ kG. 
This behavior can be linked to the crossover from separated to overlapped Landau level regime.
Using the quantum scattering time $\tauq \simeq 19$ ps extracted from the Dingle analysis of HIROs we find \citep{zhang:2007a,hatke:2009c} that $\oc \tauq \simeq 5$ at $B\simeq 1$ kG which suggests that the ZdRS form in separated Landau levels.

Further examination of $T=1.5$ K data in Fig.\,\ref{f3} reveals that the ZdRS becomes wider at higher magnetic fields as its higher critical current $I_2$ increases with $B$.
At $I > I_2$ the differential resistivity grows and then shows a fundamental ($\edc \simeq 1$) HIRO peak (cf., $\uparrow $) which occurs at $I_H = j_H\cdot w$, where $j_H=e n_e (\oc/2k_F) \propto B$ and $k_F=\sqrt{2\pi n_e}$ is the Fermi wave number.
Therefore, the increase of $I_{2}$ is largely determined by the increase of $I_H$ and thus is related to the onset of inter-Landau level scattering.
We proceed by fitting the experimental data with $r_{xx}(I)=r_{xx}(I_H)\exp[-(I-I_H)^2/\Delta_2^2]$ and find that, similar to the width of the zero-bias peak $\Delta_1$, $\Delta_2$ is roughly $B$ independent.
It is, however, noticeably larger, ranging from $\simeq 18$ to $\simeq 22$ $\mu$A.
A rough estimate for $I_2$ can be obtained as $I_2 \simeq I_H-2\Delta_2$.

%%%%%%%%%%%%%%%%%%%%%%%%%%%%%%%%%%%%%%%%%%%%%%%%%
\begin{figure}[t]
\includegraphics{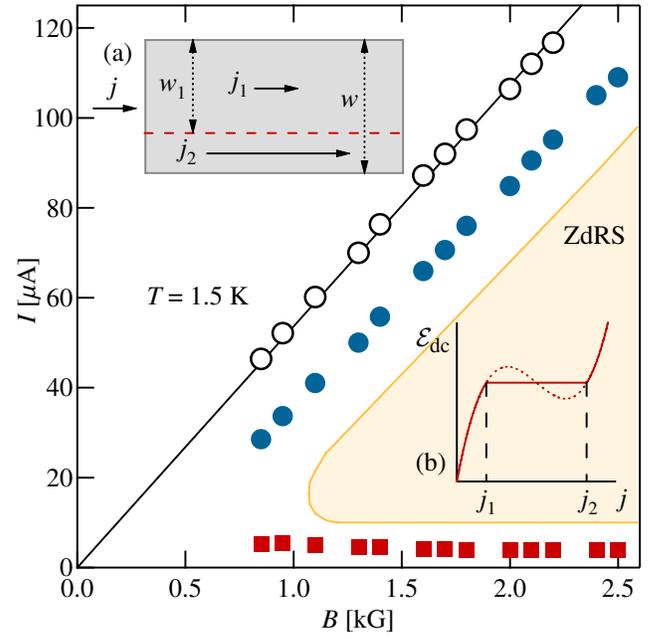}
\vspace{-0.1 in}
\caption{(Color online)
Current at the fundamental HIRO peak $I_H$ (open circles), $\Delta_1$ (squares), and $I_H-\Delta_2$ (solid circles) vs $B$. 
Shaded area marks the region $I_1(B)\lesssim I \lesssim I_2(B)$ where the ZdRS is formed.
Inset (a) shows the simplest domain structure containing domains of width $w_1$, current $j_1$ (top) and of width $w_2=w-w_1$, current $j_2$ (bottom) separated by a wall (dashed line).
Inset (b) depicts a generic $\Edc$ vs $j$ dependence with domain currents $j_1$ and $j_2$. 
}
\vspace{-0.1 in}
\label{f4}
\end{figure}
%%%%%%%%%%%%%%%%%%%%%%%%%%%%%%%%%%%%%%%%%%%%%%%%%

To summarize our experimental observations we construct a ``phase diagram'' in the $(I,B)$ plane which is presented in Fig.\,\ref{f4}.
We observe that the experimental position of the fundamental HIRO peak $I_H$ (cf., open circles) is well described by a linear relation (cf., solid line) computed using $\edc = 1$.
The extracted $\Delta_1$ and $I_H-\Delta_2$ are shown by solid squares and circles, respectively.
The shaded area roughly marks the phase space, $I_1(B)\lesssim I \lesssim I_2(B)$ where the ZdRS is formed. 
In the simplest case of two domains\citep{bykov:2007} these currents should be associated with the currents inside the domains, $I_i=j_i \cdot w_i\,\,(i=1,2)$, where $j_1$ is the domain current density and $w_i$ is the domain width [see insets (a) and (b)].
The position of the domain wall can be found from the boundary condition $I=I_1+I_2$ as $w_1/w=(I_2-I)/(I_2-I_1)$.
For $B=1.5$ kG, we estimate $I_1\simeq 10$ $\mu$A, $I_2\simeq 43$ $\mu$A, and for $I=20$ $\mu$A obtain $w_1/w \simeq 23/33 \simeq 0.7$, the situation depicted in the inset (a).
As the current approaches either $I_1$ or $I_2$, the domain wall moves to the sample boundary and the ZdRS is destroyed.\citep{bykov:2007}

In summary, we reported on a state with a zero differential resistance in a dc-driven high-mobility 2DES subject to weak magnetic fields and low temperatures.
This state emerges from a minimum of Hall field-induced resistance oscillations in the absence of microwave radiation and disappears in the regime of strongly overlapped Landau levels and with increasing temperature.
Occurring in very high Landau levels, the state extends over a continuous range of electric and magnetic fields persisting far below the onset of the Shubnikov-de Haas oscillations.
The minimum current required to support this state is largely independent on the magnetic field and the maximum current traces the onset of inter-Landau level scattering increasing linearly with the magnetic field.
According to the domain model\citep{bykov:2007} these currents should be associated with currents inside the domains formed in a dc-driven high-mobility 2DES.
To explain the temperature dependence and the mechanism leading to the ZdRS, theories might need to consider the effects of electron-electron and electron-phonon scattering, and be extended to the regime of separated Landau levels.
Since the state under study is similar to a microwave-induced zero-resistance state in a sense that it can also be explained by the domain model, it offers exciting experimental opportunities.
In particular, it might allow one to explore instabilities leading to domain formation.
Such studies have proven difficult in irradiated 2DES and no direct experimental confirmation of domains is currently available.

We thank I. A. Dmitriev and B. I. Shklovskii for useful discussions and remarks.
The work at Minnesota was supported by the NSF Grant No. DMR-0548014. 

%\vspace{-0.25in}

\end{document}